\documentclass[]{pasj01}
\draft

\usepackage{lscape}
\usepackage{lineno}

\newcommand{\swift}{\textit{Swift}}

\newcommand{\fermi}{\textit{Fermi}-LAT}

\newcommand{\src}{S4~0954+65}

\newcommand{\cm}{\mathrm{cm}}

\newcommand{\erg}{\mathrm{erg}}

\newcommand{\s}{\mathrm{s}}

\newcommand{\e}{\epsilon}

\newcommand{\psim}{\lower.5ex\hbox{$\; \buildrel \propto \over\sim \;$}}
\newcommand{\lbar}{\lower.0ex\hbox{$\; \buildrel
{\lower0.0ex \hbox{-}} \over\lambda  \;$}}

\begin{document} 
\Received{}%{yyyy/mm/dd}
\Accepted{2016/04/12}%{yyyy/mm/dd}
%\Published{yyyy/mm/dd}

\title{A significant hardening and rising shape detected in the MeV/GeV $\nu F_{\nu}$ spectrum from the recently-discovered very-high-energy blazar \src\ during the bright optical flare in 2015 February}

%%% begin:list of authors
% Do NOT capitalize all letters in "textsc".
\author{Y.~T. \textsc{Tanaka}\altaffilmark{1}, J.~\textsc{Becerra Gonzalez}\altaffilmark{2, 3}, R.~\textsc{Itoh}\altaffilmark{4}, J.~D.~\textsc{Finke}\altaffilmark{5}, Y.~\textsc{Inoue}\altaffilmark{6}, R.~\textsc{Ojha}\altaffilmark{2, 7, 8}, B.~\textsc{Carpenter}\altaffilmark{2, 8}, E.~\textsc{Lindfors}\altaffilmark{9}, F.~\textsc{Krau\ss}\altaffilmark{10, 11}, R.~\textsc{Desiante}\altaffilmark{12, 13}, K.~\textsc{Shiki}\altaffilmark{4}, Y.~\textsc{Fukazawa}\altaffilmark{4}, F.~\textsc{Longo}\altaffilmark{14, 15}, J.~\textsc{McEnery}\altaffilmark{2, 3}, S.~\textsc{Buson}\altaffilmark{2, 7}, K.~\textsc{Nilsson}\altaffilmark{16}, V.~\textsc{Fallah Ramazani}\altaffilmark{9}, R.~\textsc{Reinthal}\altaffilmark{9}, L.~\textsc{Takalo}\altaffilmark{9}, T.~\textsc{Pursimo}\altaffilmark{17}, W.~\textsc{Boschin}\altaffilmark{18, 19, 20}
}%

\email{ytanaka@hep01.hepl.hiroshima-u.ac.jp}

\altaffiltext{1}{Hiroshima Astrophysical Science Center, Hiroshima University, 1-3-1 Kagamiyama, Higashi-Hiroshima 739-8526, Japan}
\altaffiltext{2}{NASA Goddard Space Flight Center, 8800 Greenbelt Rd, Greenbelt, MD 20771, USA}
\altaffiltext{3}{Department of Physics and Department of Astronomy, University of Maryland, College Park, MD 20742, USA}
\altaffiltext{4}{Department of Physical Sciences, Hiroshima University, Higashi-Hiroshima, Hiroshima 739-8526, Japan}
\altaffiltext{5}{Space Science Division, Naval Research Laboratory, Washington, DC 20375-5352, USA}
\altaffiltext{6}{Institute of Space and Astronautical Science, JAXA, 3-1-1 Yoshinodai, Chuo-ku, Sagamihara, Kanagawa 252-5210, Japan}
\altaffiltext{7}{University of Maryland, Baltimore County, 1000 Hilltop Circle, Baltimore, MD 21250, USA}
\altaffiltext{8}{The Catholic University of America, 620 Michigan Ave NE, Washington, DC 20064}
\altaffiltext{9}{Tuorla Observatory, Department of Physics and Astronomy, University of Turku, Finland}
\altaffiltext{10}{Dr. Remeis Sternwarte \& ECAP, Universit\"{a}t Erlangen-N\"{u}rnberg, Sternwartstrasse 7, 96049 Bamberg, Germany}
\altaffiltext{11}{Institut f\"{u}r Theoretische Physik und Astrophysik, Universit\"{a}t W\"{u}rzburg, Emil-Fischer-Str. 31, 97074 W\"{u}rzburg, Germany}
\altaffiltext{12}{Istituto Nazionale di Fisica Nucleare, Sezione di Torino, I-10125 Torino, Italy}
\altaffiltext{13}{Universit\`a di Udine, I-33100 Udine, Italy}
\altaffiltext{14}{Istituto Nazionale di Fisica Nucleare, Sezione di Trieste, I-34127 Trieste, Italy}
\altaffiltext{15}{Dipartimento di Fisica, Universit\`a di Trieste, I-34127 Trieste, Italy}
\altaffiltext{16}{Finnish Centre for Astronomy with ESO, University of Turku, Finland}
\altaffiltext{17}{Nordic Optical Telescope, Apartado 474, E-38700 Santa Cruz de La Palma, Santa Cruz de Tenerife, Spain}
\altaffiltext{18}{Fundaci\'on G. Galilei-INAF (Telescopio Nazionale Galileo), Rambla J. A. Fern\'andez P\'erez 7, E-38712 Bre\~na Baja (La Palma), Spain}
\altaffiltext{19}{Instituto de Astrof\'isica de Canarias, C/V\'ia L\'actea s/n, E-38205 La Laguna (Tenerife), Spain}
\altaffiltext{20}{Departamento de Astrof\'isica, Univ. de La Laguna, Av. del Astrof\'isico F. S\'anchez s/n, E-38205 La Laguna (Tenerife), Spain}

%% `\KeyWords{}' always has to be placed before `\maketitle'.
 %Do NOT move this preamble from here!

\maketitle

\begin{abstract}
We report on \textit{Fermi} Large Area Telescope (LAT) and multi-wavelength results on the recently-discovered very-high-energy (VHE, $E>100$~GeV) blazar \src\ ($z=0.368$) during an exceptionally bright optical flare in 2015 February. During the time period (2015 February, 13/14, or MJD~57067) when the MAGIC telescope detected VHE $\gamma$-ray emission from the source, the \fermi\ data indicated a significant spectral hardening at GeV energies, with a power-law photon index of $1.8 \pm 0.1$---compared with the 3FGL value (averaged over four years of observation) of $2.34 \pm 0.04$. In contrast, \swift/XRT data showed a softening of the X-ray spectrum, with a photon index of $1.72 \pm 0.08$ (compared with $1.38\pm0.03$ averaged during the flare from MJD~57066 to 57077), possibly indicating a modest contribution of synchrotron photons by the highest-energy electrons superposed on the inverse Compton component. Fitting of the quasi-simultaneous ($<1$ day) broadband spectrum with a one-zone synchrotron plus inverse-Compton model revealed that GeV/TeV emission could be produced by inverse-Compton scattering of external photons from the dust torus. We emphasize that a flaring blazar showing high flux of $ \gtrsim 1.0 \times 10^{-6}$ photons cm$^{-2}$ s$^{-1}$ ($E>100$~MeV) and a hard spectral index of $\Gamma_{\rm GeV} < 2.0$ detected by \fermi\ on daily time scales is a promising target for TeV follow-up by ground-based Cherenkov telescopes to discover high-redshift blazars, investigate their temporal variability and spectral features in the VHE band, and also constrain the intensity of the extragalactic background light.
\end{abstract}

%\linenumbers

\section{Introduction}
\label{sec-intro}
In the diverse family of active galactic nuclei (AGN), blazars stand out due to their extreme variability in all wavebands and over a broad range of timescales. Their predominantly non-thermal emission arises in relativistic jets that are pointed close to our line of sight. The resulting Doppler boosting is responsible for their short-timescale variability, apart from boosting their flux and creating the illusion of superluminal motion (e.g., \cite{Urry95}). This broadband variability presents both a challenge and an opportunity. On the one hand, the variability makes it difficult to construct a physical model of high-energy emission from blazars. On the other hand, the variability also provides important constraints on the many open questions about the origin of blazar emission. With continuous monitoring of the sky by the {\it Fermi} Gamma-ray Space Telescope, and observations by X-ray satellites as well as ground-based telescopes in the radio through TeV bands, we are able to make near-simultaneous observations that contribute to addressing these questions (e.g., \cite{FermiMrk501, FermiMrk421}).

Blazars are typically divided into BL\,Lac objects and flat spectrum radio quasars (FSRQs) with the formal distinction being the absence or presence, respectively, of emission lines with a rest frame equivalent width $\geq 5$~\AA\ (e.g., \cite{Marcha96}). \src\ is a blazar at a redshift $z=0.368$ \citep{Stickel93, Lawrence96}. Although a recent paper by \citet{Landoni15} reported a more distant lower limit to the redshift at $z \geq 0.45$, our preliminary result for the source spectrum taken with the Telescopio Nazionale Galileo 3.58~m telescope confirm the $z=0.368$ (Becerra Gonzalez et al. in prep.). This object clearly meets the formal definition of a BL\,Lac (see Table~35 and Fig.~8 of \cite{Lawrence96}). However, its archival (non-simultaneous) multi-wavelength spectral energy distribution (SED) hints at the presence of a ``blue bump" more typical of a FSRQ. Past X-ray observation by ROSAT (e.g., \cite{Comastri97}) shows a flatter energy distribution than typical for a radio-selected BL\,Lac leading to the suggestion that \src\ may be a transition object with properties that lie in between the BL\,Lac and FSRQ classes. This idea has also been explored by \citet{Ghisellini11}, who, however, conclude that it should be classified as a LBL (a ``low-peaked" BL\,Lac object) based on the luminosity of the broad-line region in Eddington units, rather than the emission lines' equivalent width.

A powerful $\gamma$-ray flare was detected from \src\ by {\it Fermi} Large Area Telescope (LAT) on 2014 November 25 \citep{Krauss14} when its daily averaged $\gamma$-ray flux ($E>100$~MeV) was about 32 times its average flux in the \fermi\ third source catalog (3FGL catalog, see \cite{3FGL}). In late January 2015, \citet{Carrasco15a} reported an increase by a factor of three in its near-infrared (NIR) emission. This heralded the beginning of unprecedented optical/NIR activity in this object with its $V$-band magnitude brightening by two magnitudes \citep{Stanek15}, continued flaring in the NIR band \citep{Carrasco15b}, and its brightest ever optical state reported \citep{Spiridonova15a, Spiridonova15b}. Rapid intra-night variability in the $R$-band was detected on 11-15 February 2015 \citep{Bachev15}. An increase in the degree of optical polarization in the $R$-band was also observed from 14\% on 18 February 2015 to 25\% on 19 February 2015 \citep{Jorstad15}.

On 2015 February 13/14 (MJD~57067) the MAGIC telescopes detected very-high-energy (VHE; $\mathrm{E > 100\,GeV}$) emission from \src\ \citep{Mirzoyan15}. This coincided with the detection of an unusually hard $\gamma$-ray ($E>0.1$~GeV) spectrum by \fermi\ along with an elevated $\gamma$-ray flux \citep{Ojha15}. In this paper, we make a detailed study of the evolution of the $\gamma$-ray spectrum and its relationship to activity in the X-ray and optical bands. We first present our observations in \S\ref{sec-data}. Then we show the results in \S\ref{sec-results}, and discuss them in \S\ref{sec-dis}. Throughout this paper, we use the cosmology $\mathrm{{\it H}_0 = 70 \,km\,s^{-1} Mpc^{-1}}$, $\Omega_{m}=0.3$, and $\Omega_{\Lambda} = 0.7$ \citep{Komatsu09}. Note that \src\ is listed in the second {\it Fermi}-LAT catalog of high-energy sources (2FHL catalog, see \cite{2FHL}) as 2FHL~J0958.3+6535.

\section{Observations}
\label{sec-data}

\subsection{\fermi}
\label{sec-LAT}
The LAT on board the {\it Fermi} satellite monitors the entire $\gamma$-ray sky every 3 hours in the energy range from 20~MeV to $> 300$~GeV \citep{Atwood09}. We selected Pass 7 reprocessed source-class events, from 4 August 2008 to 30 April 2015, within a 10 deg circular region centered at the location of \src. The analysis was performed with the ScienceTools software package version v9r33p0 using the instrument response function \texttt{P7REP\_SOURCE\_V15} \citep{Ackermann12}. A zenith angle cut of $< 100^{\circ}$ was applied to reduce the contamination from the Earth Limb. The appropriate Galactic diffuse emission model (\texttt{gll\_iem\_v05\_rev.fit}) and isotropic component (\texttt{iso\_source\_v05.txt}) were used\footnote{Available at http://fermi.gsfc.nasa.gov/ssc/data/access/lat/BackgroundModels.html}. The normalizations of both components in the background model were allowed to vary freely during the spectral fitting. The unbinned maximum-likelihood method implemented in the \texttt{gtlike} tool was used. For a first likelihood fit, the model included all the 3FGL \citep{3FGL} sources within a 15$^{\circ}$ circular region around \src. Spectral indices and fluxes were left free for the fit for sources within 10$^{\circ}$, while sources from 10$^{\circ}$ to 15$^{\circ}$ were frozen to the catalog values. The significance of each source was evaluated using the test statistic ${\rm TS} = 2 \left( \log L_1 - \log L_0 \right)$, where $L$ is the likelihood of the data given the model with ($L_1$) or without ($L_0$) the source and TS is interpreted as a detection significance of $\sim \sqrt{\rm TS} \sigma$ (e.g., \cite{Mattox96}). A maximum-likelihood analysis was performed with several iterations to remove sources not contributing to the Region of Interest (low TS values, up to a maximum of ${\rm TS}=10$). The light curve has been calculated in 30, 7, and 1-day time bins modeling the source with a single power-law spectrum (as described in the 3FGL catalog). Both the flux and spectral index of \src\ were left free during the light curve calculation, while the rest of the point sources were fixed and only the diffuse Galactic and isotropic models were allowed to vary.

The LAT SEDs were calculated for four time intervals which show different characteristics in the multi-wavelength light curve (see \S~\ref{sec-results} for details). In all cases the spectrum is well-fit by a single power law (PL). A curvature test was performed on the SEDs in each time interval assuming a log-parabolic (LP) fit for comparison with the power law. As defined in \citet{Nolan12}, the curvature test statistic can be expressed as ${\rm TS}_{\rm curve}=\left( {\rm TS}_{\rm LP} - {\rm TS}_{\rm PL} \right)$. We do not find significant curvature in any of the above periods.

\subsection{X-ray}
\label{sec-xray}

The {\it Swift} X-Ray Telescope (XRT, \cite{Burrows05}) observed \src\ many times since July 2006, and all the XRT data presented here were taken in photon counting (PC) mode. Data reduction and calibration were performed with HEASoft v6.4 standard tools. We selected events of 0.3--8~keV and grades 0--12 for analysis. Source spectra were binned to include a minimum of 20 counts in each bin to allow $\chi^2$ minimization fitting. Response files were generated with \texttt{xrtmkarf}, with corrections applied for point-spread function losses and CCD defects. For spectral analysis we used the \verb|XSPEC| software package version 12.3.0.

We fit the \swift/XRT data by assuming an absorbed single power-law model where hydrogen column density for the direction of \src\ is fixed to the Galactic value of $N_H = 4.8 \times 10^{20}$ cm$^{-2}$, which is estimated from the Leiden/Argentine/Bonn (LAB) Survey of Galactic HI \citep{Kalberla05}. All the data were well represented by the absorbed power-law model except that taken on MJD~57077 (obsID: 00033530018) for which a broken power-law model is applied (see \S\ref{sec-results} for details).

\subsection{Optical and ultraviolet photometry}
\label{sec-optical}

We analyzed optical and ultraviolet data in $V$, $B$, $U$, $UVW1$, $UVM2$ and $UVW2$ bands taken with the Ultraviolet and Optical Telescope (UVOT, \cite{Roming05}) onboard \swift. The UVOT data were reduced following the standard procedure for CCD photometry. Source counts were extracted from a circular region of 5 arcsec radius, while background counts were measured from an annulus centered on the target position with inner and outer radii of 27.5 and 35 arcsec, respectively. The net source counts were converted to flux densities using the standard zero points \citep{Poole08}. The fluxes were corrected for Galactic extinction \citep{Schlegel98} to obtain the intrinsic fluxes ($A_{\rm V}=0.321, A_{\rm B}=0.436, A_{\rm U}=0.492, A_{\rm UVW1}=0.784, A_{\rm UVM2}=1.146, A_{\rm UVW2}=1.091$).

The source was observed in the optical R-band as part of the Tuorla blazar monitoring program\footnote{http://users.utu.fi/kani/1m} \citep{Takalo08}. These observations were made using the 35~cm Celestron telescope attached to the KVA 60~cm telescope (La Palma, Canary Islands, Spain). The data have been analyzed using the semi-automatic pipeline developed at the Tuorla Observatory (Nilsson et al. 2016, in prep.). The observed fluxes have been corrected for Galactic extinction using values from \citet{Schlafly11} (see Appendix of their paper). \src\ was also observed by 2.56~m Nordic Optical Telescope (NOT) in SDSS (Sloan Digital Sky Survey) $u$ and $z$ bands. The data were reduced (de-biasing, flat field correction) using standard IRAF routines. By using aperture photometry with the typical aperture radius $1.0-1.5$ arcsec, we measured the source magnitudes against the stars 3 and 6 in \citet{Raiteri99}. 

\section{Results}
\label{sec-results}
\begin{figure}[!th]
\begin{center}
\includegraphics[width=14cm]{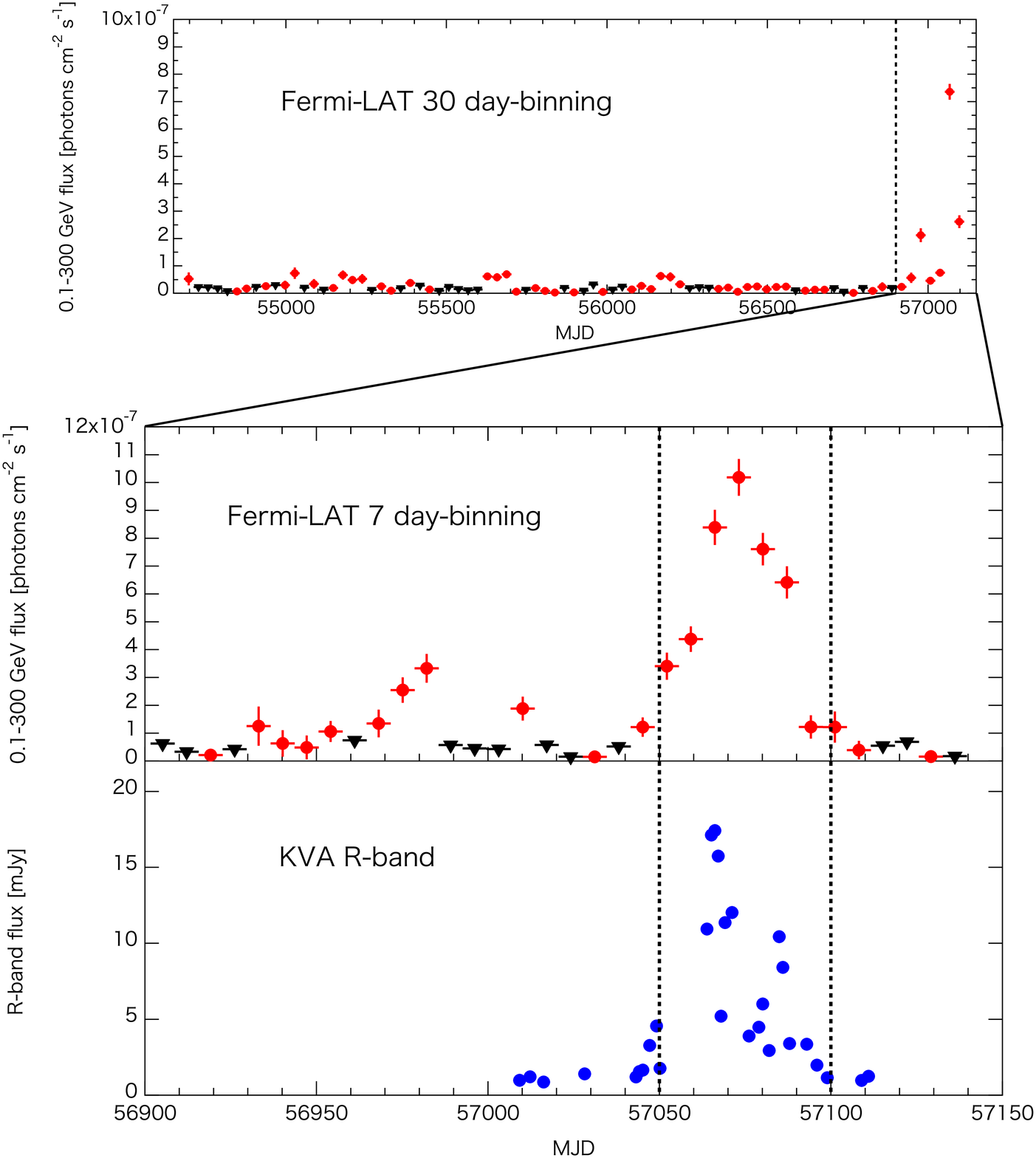}
\caption{
({\it Upper}) \fermi\ 30-day binned 0.1--300~GeV flux light curve of \src\ from 2008 August to 2015 April. Black triangles show 90\% confidence level upper limits when ${\rm TS} < 4$. They are calculated by assuming a single power-law spectrum of $\Gamma=2.34$, taken from 3FGL catalog \citep{3FGL}. ({\it Lower}) \fermi\ 7-day binned and KVA daily $R$-band extinction-corrected ($A_{\rm R}=0.259$) light curves during high state from MJD~56900 to 57150. The two vertical dashed lines indicate the period of ``highest" state from MJD~57050 to 57100. Note that daily light curves during the ``highest" state in $\gamma$-ray, X-ray, optical and UV bands are shown in Fig.~\ref{fig:lc2}.
}
\label{fig:lc}
\end{center}
\end{figure}

\begin{figure}[!th]
\begin{center}
\includegraphics[width=14cm]{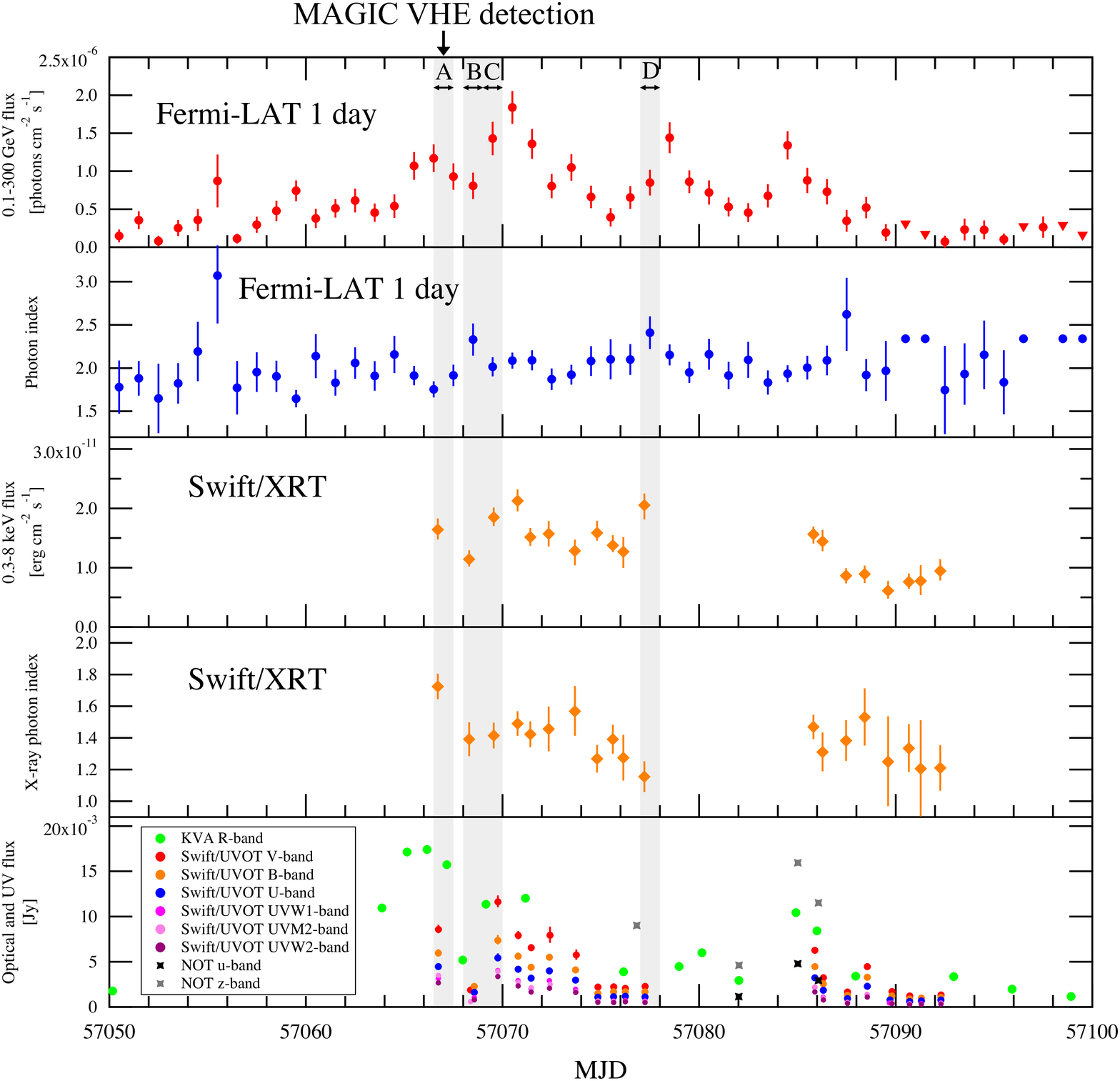}
\caption{Multi-wavelength light curves of \src\ during the ``highest" state between MJD~57050 and 57100. From {\it top} to {\it bottom}: \fermi\ 1-day binned 0.1--300~GeV flux, \fermi\ daily photon power-law index, \swift/XRT flux (0.3--8~keV), \swift/XRT photon power-law index, and optical/UV fluxes in 7 bands measured by KVA and \swift/UVOT.
Gray hatched areas, labeled by A, B, C, and D, indicate the selected 1-day periods during which SEDs are constructed (see Fig.\ref{fig:sed}). The black arrow at the top indicates the time (MJD~57067.0) when MAGIC telescope detected VHE emission \citep{Mirzoyan15}. In the second panel, the blue points with no error indicate the 3FGL value of 2.34, which was assumed for flux upper limit calculation.
}
\label{fig:lc2}
\end{center}
\end{figure}

Figure~\ref{fig:lc} displays the \fermi\ 30-day binned light curve from 2008 August to 2015 April. \src\ entered a high state after MJD~56900 and hence we produced a \fermi\ weekly (7-day) binned light curve together with a daily KVA R-band one during the high state (Fig.~\ref{fig:lc}, lower panel). The brightening in the $\gamma$-ray and optical bands is prominent in particular between MJD 57050 and 57100. To investigate the details of flux and spectral changes in multiple bands, we constructed \fermi, \swift/XRT, \swift/UVOT, and KVA light curves in each of the time periods and they are shown in Fig.~\ref{fig:lc2}. Note that the MAGIC telescope detected sub-TeV emission on MJD~57067.0 \citep{Mirzoyan15}. Indeed, on MJD~57066 and 57067, \fermi\ detected a moderate 0.1--300~GeV flux of $\sim 1.0 \times 10^{-6}$ photons cm$^{-2}$ s$^{-1}$ but with an unusually hard spectrum of $\Gamma_{\rm GeV} < 2.0$, where $\Gamma_{\rm GeV}$ is the photon power-law index on daily time scales in the LAT band (see second panel of Fig.~\ref{fig:lc2}). Note here that the 4-year averaged power-law index of the LAT spectrum is $2.38\pm0.04$ \citep{3FGL} and that a similarly hard GeV spectrum was observed on MJD~57059. 

% LAT and MWL light curves in 2015 Feb

Interestingly, the quasi-simultaneous ($< 1$ day) \swift/XRT spectrum showed a clear softening ($\Gamma_{\rm x}=1.72\pm0.08$) compared to that measured on the other days during the high state shown here ($\Gamma_{\rm x}=1.38\pm0.03$, see Table~\ref{tab:xrt}). The simultaneous R-band flux was almost at the brightest level during this outburst. 

Note also that \fermi\ detected a 51~GeV photon from close vicinity of \src\ on MJD 57066.98, which was exactly simultaneous with the time of the MAGIC VHE detection. The angular separation between this 51 GeV event and the position of \src\ was only 0.013$^{\circ}$ and the probability that the event belongs to \src\ was $>99$\% based on the \texttt{gtsrcprob} tool available in the ScienceTools. The quasi-simultaneous SED on MJD 57066.5--57067.5 (period A), which is selected to include the MAGIC VHE detection time, is shown in the upper-left panel in Fig.~\ref{fig:sed}.

On the next day (MJD~57068--57069, period B), the 0.1--300~GeV flux slightly decreased and the LAT spectrum became softer ($\Gamma=2.3 \pm 0.2$), while the X-ray spectrum became harder. In addition, the optical flux showed a sharp decrease. On MJD~57069--57070 (period C), GeV $\gamma$-ray, X-ray and optical fluxes increased again. The \fermi\ and \swift/XRT spectra were intermediate with power-law indices of $\Gamma_{\rm GeV}=2.0\pm0.1$ and $\Gamma_{\rm x}=1.41\pm0.08$, respectively. After that, fluxes in the MeV/GeV, X-ray, and optical bands showed a gradual decrease with an almost similar spectral shape, but on MJD~57077--57078 (period D), the X-ray spectrum showed the hardest index during this outburst. Note here that the limited statistics of \fermi\ makes it hard to draw strong conclusions on the evolution of the $\gamma$-ray spectral index between period B and D. We checked the XRT data on Period D and found that larger systematic residuals are present in the lower and higher energy and hence we fitted the data using a broken power-law model. The broken power-law model is statistically favored over a single power law (p-value of $5.1 \times 10^{-4}$ from an $F$-test). The best-fit values were $\Gamma_{\rm low}=0.78^{+0.21}_{-0.22}$, $\Gamma_{\rm high}=1.90^{+0.57}_{-0.39}$, and $E_{\rm break}=2.66^{+0.70}_{-0.48}$~keV. Note that \citet{Ghisellini11} also claimed from \swift/XRT data accumulated over 2006 to 2010 that a broken power law is a better representation for the X-ray spectrum of \src\ (see Table~2 of their paper).

% Swift/XRT spectral parameters
\begin{table}
\caption{\swift/XRT power-law indices and fluxes during MJD~57066--57078, the GeV-brightest period}
\begin{center}
\label{tab:xrt}
\begin{tabular}{ccc}
\hline
MJD & PL index & 0.3--8~keV flux \\
 & & ($10^{-11}$ erg cm$^{-2}$ s$^{-1}$) \\
\hline
57066.71  &  1.72$\pm$0.08     & 1.64$\pm$0.19 \\
57068.30  &  1.39$\pm$0.11    & 1.14$\pm$0.15 \\
57069.55  &  1.41$\pm$0.08    & 1.85$\pm$0.16 \\
57070.76  &  1.49$\pm$0.08    & 2.13$\pm$0.19 \\
57071.42  &  1.42$\pm$0.08    & 1.51$\pm$0.15 \\
57072.35  &  1.46$\pm$0.14     &1.57$\pm$0.21 \\
57073.68  &  1.57$\pm$0.16     &1.28$\pm$0.24 \\
57074.81  &  1.27$\pm$0.09     &1.59$\pm$0.20 \\
57075.61  &  1.39$\pm$0.09     &1.38$\pm$0.17 \\
57076.15  &  1.27$\pm$0.14     &1.27$\pm$0.27 \\
57077.21  &  1.15$\pm$0.10    & 2.05$\pm$0.24 \\
\hline
\end{tabular}
\end{center}
\end{table}

\section{Discussion}
\label{sec-dis}
\begin{figure}[!th] 
\begin{center}
\includegraphics[width=\textwidth]{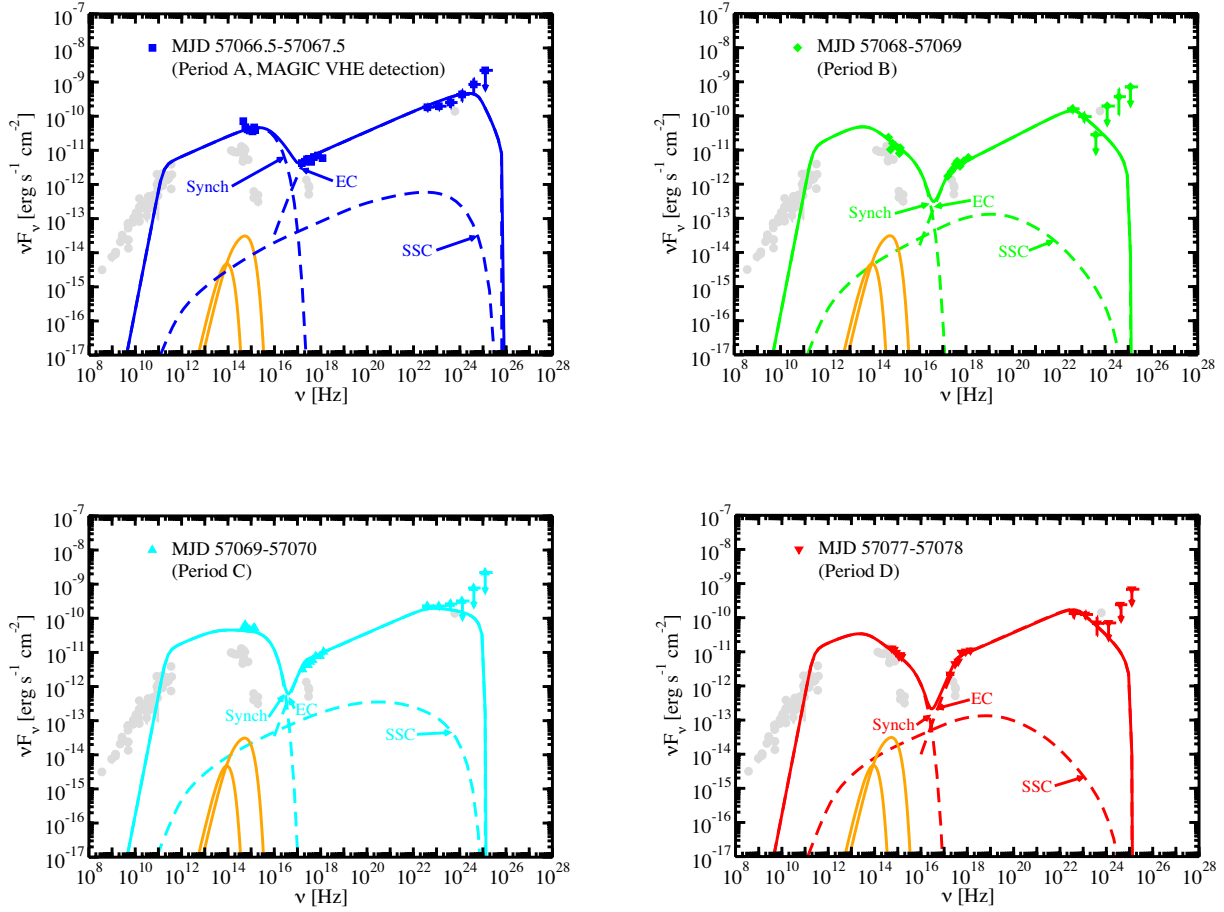}
\caption{Quasi-simultaneous ($<1$~day) SEDs of \src\ during the 4 selected time intervals. ({\it Top-left}) The SED on MJD~57066.5--57067.5, including MAGIC VHE detection time. Optical/UV data are taken from KVA R-band and \swift/UVOT measurements. X-ray and MeV/GeV fluxes are from \swift/XRT and \fermi, respectively. The blue line indicates a model curve (Synchrotron, EC, and SSC emissions are summed up) calculated based on one-zone synchrotron emission and inverse-Compton scattering of dust torus photons \citep{Finke08, Dermer09}. The two orange lines indicate the dust torus and accretion disk emissions. The gray circles are historical fluxes taken from the NED database. The derived parameter values are tabulated in Table~\ref{tab:sed}.
({\it Top-right}) Same as {\it top-left} panel but for the SED on MJD~57068--57069 (shown in green). ({\it Bottom-left}) Same as {\it top-left} panel but for the SED on MJD~57069--57070 (shown in cyan). ({\it Bottom-right}) Same as {\it top-left} panel but for the SED on MJD~57077--57078 (shown in red). KVA R-band flux is not included during this period due to lack of observation.
}
\label{fig:sed}
\end{center}
\end{figure}

To derive physical quantities at the emission site, the broadband spectra for the four selected periods are modeled by a one-zone synchrotron plus inverse-Compton model \citep{Finke08, Dermer09}. The electron distribution is assumed to have a broken power-law shape,
\begin{eqnarray*}
N^{\prime} \left( \gamma^{\prime} \right) &\propto& \gamma^{\prime -s_1} \quad \left( \gamma^{\prime}_{\rm min} < \gamma^{\prime} < \gamma^{\prime}_{\rm brk} \right)\\
N^{\prime} \left( \gamma^{\prime} \right) &\propto& \gamma^{\prime -s_2} \quad \left( \gamma^{\prime}_{\rm brk} < \gamma^{\prime} < \gamma^{\prime}_{\rm max} \right),
\end{eqnarray*}
where $\gamma^{\prime}_{\rm min}$, $\gamma^{\prime}_{\rm max}$, and $\gamma^{\prime}_{\rm brk}$ are the minimum, maximum, and break electron Lorentz factors, respectively. $s_1$ and $s_2$ are the power-law indices of the electron distribution below and above the break electron Lorentz factor $\gamma^{\prime}_{\rm brk}$. 
Primed quantities indicate those measured in the jet comoving frame.
The model curves and derived parameter values are shown in Fig.~\ref{fig:sed} and Table~\ref{tab:sed}, respectively. The SEDs were well represented by changing only the electron distribution and the magnetic field (see also e.g., \cite{Dutka13, Ackermann14}). 
Note that the spectral break in the electron distribution cannot be understood in terms of radiative cooling, because $s_2-s_1$ does not correspond to the canonical value of 1.0 (e.g., \cite{Longair11}).
We found that the $\gamma$ rays can be modeled by an external Compton (EC) component, rather than synchrotron self-Compton (SSC), despite the BL Lac classification for this object \citep{Mukherjee95}. We modeled the seed photon source for this process as a monochromatic isotropic external radiation field with energy density $u_{\rm seed}=2.4\times10^{-4}$\ erg s$^{-1}$ and energy $\e_0=7.5\times10^{-7}$ in $m_ec^2$ units. This corresponds to a dust temperature of $T_{\rm dust} = 1500$\ K and, for a disk luminosity of $3.0\times10^{43}$\ erg s$^{-1}$  and, using the relation from \citep[equation (1)]{Nenkova08}, a dust radius of $2.1\times10^{17}$\ cm.
Note that, as shown in Fig.~\ref{fig:sed}, the SSC component is lower than the EC one by two orders of magnitude under the parameter values tabulated in Table~\ref{tab:sed}.
Note also that once we assume that SSC emission is responsible for the X-ray and MeV/GeV $\gamma$-ray emissions, the required magnetic field becomes very small ($B \sim 1$~mG) because of the relatively large Compton dominance of $L_{\rm IC}/L_{\rm sync} \sim 10$. Since this is much weaker than the typical magnetic field derived from  blazar SED modeling ($\sim 1$~Gauss, see e.g., \citet{Ghisellini10}), our modeling under the EC assumption seems reasonable. There would be another option that the X-ray and MeV/GeV emissions are from SSC and EC components, respectively. However, given the lack of evidence of a spectral break between the X-ray and MeV/GeV data points, it is simpler to assume that only a single EC component is responsible for both X-ray and MeV/GeV emissions. In this regard, more precise flux measurements are needed to determine whether our assumption is valid or an alternative SSC+EC modeling is required.

% SED and modeling

% SED parameter table
%\begin{landscape}
\begin{table}
\scriptsize
\caption{Model parameters.}
\begin{center}
\label{tab:sed}
\begin{tabular}{lccccc}
%\rotate
\hline
Parameter & Symbol & MJD 57066.5-57067.5 & MJD 57068-57069 & MJD 57069-57070 & MJD 57077-57078 \\
\hline
Redshift & 	$z$	& \multicolumn{4}{c}{0.368} \\
Bulk Lorentz Factor & $\Gamma$	& \multicolumn{4}{c}{30} \\
Doppler factor & $\delta_D$	& \multicolumn{4}{c}{30} \\
Variability Timescale [s]& $t_v$       & \multicolumn{4}{c}{$1.0\times10^5$} \\
Comoving radius of blob [cm]& $R^{\prime}_b$ & \multicolumn{4}{c}{6.6$\times$10$^{16}$} \\
Magnetic Field [G]& $B$         & 0.6 & 1.4 & 1.0 & 1.0 \\
\hline
Low-Energy Electron Spectral Index & $s_1$   & 2.4 & 2.3 & 2.4 & 2.4 \\
High-Energy Electron Spectral Index  & $s_2$ & 4.5 & 4.0 & 3.0 & 4.0 \\
Minimum Electron Lorentz Factor & $\gamma^{\prime}_{\rm min}$  & $1.0$ & $1.0$ & $1.0$ & $1.5$ \\
Break Electron Lorentz Factor & $\gamma^{\prime}_{\rm brk}$ & $8.0\times10^3$ & $6.0\times10^2$ & $6.0\times10^2$ & $6.0\times10^2$ \\
Maximum Electron Lorentz Factor & $\gamma^{\prime}_{\rm max}$  & $2.0\times10^4$ & $1.0\times10^4$ & $1.0\times10^4$ & $1.0\times10^4$ \\
\hline
Black hole Mass [$M_\odot]$ & $M_{\rm BH}$ & \multicolumn{4}{c}{$3.4\times10^8$} \\
Disk luminosity [$\erg\ \s^{-1}$] & $L_{\rm disk}$ & \multicolumn{4}{c}{$3.0\times10^{43}$} \\
Inner disk radius [$R_g$] & $R_{\rm in}$ & \multicolumn{4}{c}{$6.0$}\\
Seed photon source energy density [$\erg\ \cm^{-3}$] & $u_{\rm seed}$ & \multicolumn{4}{c}{$2.4\times10^{-4}$} \\
Seed photon source photon energy [$m_e c^2$ units] & $\e_{\rm seed}$ & \multicolumn{4}{c}{$7.5\times10^{-7}$} \\
Dust Torus luminosity [$\erg\ \s^{-1}$] & $L_{\rm dust}$ & \multicolumn{4}{c}{$3.9\times10^{42}$} \\
Dust Torus radius [cm] & $R_{\rm dust}$ & \multicolumn{4}{c}{$2.1\times10^{17}$} \\
Dust temperature [K] & $T_{\rm dust}$ & \multicolumn{4}{c}{$1500$} \\
\hline
Jet Power in Magnetic Field [$\erg\ \s^{-1}$] & $P_{j,B}$ & $1.0\times10^{46}$ & $5.7\times10^{46}$ & $2.9\times10^{46}$ & $2.9\times10^{46}$   \\
Jet Power in Electrons [$\erg\ \s^{-1}$] & $P_{j,e}$ & $1.1\times10^{45}$ & $6.1\times10^{44}$ & $1.3\times10^{45}$ & $1.1\times10^{45}$  \\
\hline
\end{tabular}
\end{center}
\end{table}
%\end{landscape}

During the GeV spectral hardening (MJD~57066.5--57067.5, period A), the break energy of the electron distribution $\gamma^{\prime}_{\rm brk}$ increased about one order of magnitude (up to $8 \times 10^3$ from $6 \times 10^2$) due to the rising shape of the LAT $\nu F_{\nu}$ spectrum, indicating a rapid injection of high-energy electrons with $\gamma^{\prime} \sim 10^3$--$10^4$. The observed softer X-ray spectrum in period A would result from the modest contribution of synchrotron photons emitted by the highest energy electrons instead of the inverse-Compton X-rays produced by the lowest energy electrons (see upper left panel of Fig.~\ref{fig:sed}). We note that the spectral break at $E_{\rm break}=2.66^{+0.70}_{-0.48}$~keV seen in period D can be modeled by setting the minimum Lorentz factor of the electron distribution to be 1.5. Note also that a similar X-ray break seems to be present in the X-ray data during Period D (MJD~57068--57069), which is again reasonably modeled by $\gamma^{\prime}_{\rm min}=1.0$ (see {\it upper right} panel in Fig.~\ref{fig:sed} and Table~\ref{tab:sed}). Therefore, we stress that X-ray spectroscopy is a powerful tool to constrain the minimum electron Lorentz factor $\gamma^{\prime}_{\rm min}$ of the emitting electron distribution (see also e.g., \citet{Celotti08}). We also point out that the observed spectral break is a good indication that the EC component indeed dominates over SSC in the X-ray band, because it is difficult to produce such a break by assuming SSC.

From SED modeling, we also found that the jet power in the magnetic field ($P_{\rm B}$) dominates over the jet power in emitting electrons ($P_{\rm e}$) by a factor of 10--100 (see Table~\ref{tab:sed}). Here we define the jet power components as in \citet{Finke08}; $P_{\rm i}=2 \pi R^{\prime 2} \Gamma^2 \beta c U^{\prime}_{\rm i}$ (${\rm i}={\rm B}, {\rm e}$), where $\Gamma=\left( 1- \beta^2 \right)^{-1/2}$ is the bulk Lorentz factor of the emitting blob, $U^{\prime}_{\rm B}=B^2/8\pi$ and $U^{\prime}_{\rm e}= \left( m_e c^2/V^{\prime} \right) \int^{\gamma^{\prime}_{\rm max}}_{\gamma^{\prime}_{\rm min}} \gamma^{\prime} N^{\prime}_e \left( \gamma^{\prime} \right)$ are the energy densities of magnetic field and electrons, respectively, and $V^{\prime}= \left( 4/3 \right) \pi R^{\prime 3}$ is the volume of the emitting blob. Note that this definition assumes a two-sided jet. 
This Poynting-flux dominance is robust under our EC assumption and not unprecedented considering there are several blazars showing a similar feature of $P_B > 10 P_e$ such as 0234+285 and 0528+134 (see Table~A2 of \citet{Celotti08}).
There is some evidence that cold protons in the jet ($P_{\rm p}= 2 \pi R^{\prime 2} \Gamma^2 \beta c \left( m_p c^2/V^{\prime} \right) \int^{\gamma^{\prime}_{\rm max}}_{\gamma^{\prime}_{\rm min}} N^{\prime}_p \left( \gamma^{\prime} \right)$, where $N^{\prime}_p$ is a proton distribution and $N^{\prime}_p=N^{\prime}_e$ is assumed, see e.g., \cite{Ghisellini14}) can carry much larger (as large as 100 times) power than the emitting electrons (e.g., \cite{Sikora00, Ghisellini14, Tanaka15}). Hence, it is possible in the context of the models presented here, that $P_{\rm B} \sim P_{\rm e}+P_{\rm p}$.

This paper serves as a case study for the capability of detecting new VHE sources based upon follow-up of flaring LAT sources showing spectral hardening (i.e. fluxes above $1.0 \times 10^{-6}$ photons cm$^{-2}$ s$^{-1}$  above 100~MeV and $\Gamma_{\rm GeV} < 2.0$). The capabilities of the LAT (specifically the daily all-sky monitoring and the improved high-energy performance from Pass~8 \citep{Pass8}) are well suited to these types of efforts and we can expect many such discoveries in the next few years. In fact, several spectral hardening events have been seen from Fermi-LAT FSRQs (e.g., \cite{Tanaka11, Pacciani14}) which would have been excellent candidates for VHE follow-up at the time.  

Additionally, recent theoretical and observational studies of the extragalactic background light (EBL) indicate that the horizon of 100 GeV photons is $z \sim 1$ (e.g., \cite{Finke10, Dominguez11, Ackermann12EBL, Inoue13}).  The current capabilities of the LAT are allowing us to probe beyond this edge. For example, \citet{Tanaka13} report the detection of two VHE photons from the $z=1.1$ blazar PKS~0426-380 (see also Figure~13 of \citet{2FHL} for the \fermi\ detection of $E>50$~GeV photons from blazars beyond the horizon). But the current generation of ground based VHE observatories have not yet detected a source beyond a redshift of 1. MAGIC recently reported the detection of two high-redshift blazars S3~0218+35 at $z = 0.944$ \citep{Mirzoyan14} and PKS~1441+25 at $z = 0.939$ \citep{Mirzoyan1441, VERITAS_ApJL, MAGIC_ApJL}, but, depending on the spectrum of these sources at VHE energies, might not challenge the current understanding of the EBL. Triggering VHE observations of moderately-high redshift blazars with the \fermi\ when they are in high- and hard-flux states is a way to push the redshift limit of VHE detections further and allow us to learn more about the EBL. This will become even more important when the next generation instrument, CTA, comes online and provides a lower energy threshold combined with better sensitivity.

\begin{ack}
We appreciate the referee's careful reading and valuable comments.
The \textit{Fermi} LAT Collaboration acknowledges generous ongoing support
from a number of agencies and institutes that have supported both the
development and the operation of the LAT as well as scientific data analysis.
These include the National Aeronautics and Space Administration and the
Department of Energy in the United States, the Commissariat \`a l'Energie Atomique
and the Centre National de la Recherche Scientifique / Institut National de Physique
Nucl\'eaire et de Physique des Particules in France, the Agenzia Spaziale Italiana
and the Istituto Nazionale di Fisica Nucleare in Italy, the Ministry of Education,
Culture, Sports, Science and Technology (MEXT), High Energy Accelerator Research
Organization (KEK) and Japan Aerospace Exploration Agency (JAXA) in Japan, and
the K.~A.~Wallenberg Foundation, the Swedish Research Council and the
Swedish National Space Board in Sweden.
Additional support for science analysis during the operations phase is gratefully acknowledged from the Istituto Nazionale di Astrofisica in Italy and the Centre National d'\'Etudes Spatiales in France.
This research was funded in part by NASA through Fermi Guest Investigator grants NNH12ZDA001N and NNH13ZDA001N-FERMI. This research has made use of NASA's Astrophysics Data System. YTT is supported by Kakenhi 15K17652.
\end{ack}

\bibliographystyle{apj}
%\bibliography{bibs_0954}

\end{document}